\documentclass[review]{elsarticle}

\usepackage{lineno,hyperref}
\usepackage{graphicx}
\usepackage{booktabs}
\usepackage{color}
\usepackage{colortbl}
\usepackage{diagbox}
\usepackage{array}
\usepackage{tablefootnote}
\usepackage{amsfonts}
\usepackage{mathtools}
\usepackage{xcolor}
\usepackage{subfig}

\modulolinenumbers[5]

\journal{Journal of Information Processing and Management}

\biboptions{authoryear}

\begin{document}

\begin{frontmatter}

\title{Faceted search of heterogeneous geographic information for dynamic map projection}

\author[mymainaddress]{Noemi Mauro}
\ead{noemi.mauro@unito.it}

\author[mymainaddress]{Ardissono Liliana\corref{mycorrespondingauthor}}
\ead{liliana.ardissono@unito.it}

\author[mymainaddress]{Maurizio Lucenteforte}
\ead{maurizio.lucenteforte@unito.it}

\cortext[mycorrespondingauthor]{This is to indicate the corresponding author.}

\address[mymainaddress]{Dipartimento di Informatica, Universit\`{a} degli Studi di Torino, Corso Svizzera 185, I-10149 Torino, Italy}

\begin{abstract}
This paper proposes a faceted information exploration model that supports coarse-grained and fine-grained focusing of geographic maps by offering a graphical representation of data attributes within interactive widgets.
The proposed approach enables (i) a multi-category projection of long-lasting geographic maps, based on the proposal of efficient facets for data exploration in sparse and noisy datasets, and (ii) an interactive representation of the search context based on widgets that support data visualization, faceted exploration, category-based information hiding and transparency of results at the same time. The integration of our model with a semantic representation of geographical knowledge supports the exploration of information retrieved from heterogeneous data sources, such as Public Open Data and OpenStreetMap.

We evaluated our model with users in the OnToMap collaborative Web GIS. The experimental results show that, when working on geographic maps populated with multiple data categories, it outperforms simple category-based map projection and traditional faceted search tools, such as checkboxes, in both user performance and experience.  
\end{abstract}


\end{frontmatter}
\noindent
Declarations of interest: none.
\\
\\
\textbf{Published in Information Processing \& Management, Elsevier. 
\\
DOI: \url{https://doi.org/10.1016/j.ipm.2020.102257}. 
\\Link to the page of the paper on Elsevier web site:
\\
\url{http://www.sciencedirect.com/science/article/pii/S0306457319311987}
\\
This work is licensed under the:
\\Creative Commons Attribution-NonCommercialNoDerivatives 4.0 International License. 
\\
To view a copy of this license, visit  \url{http://creativecommons.org/licenses/by-nc-nd/4.0/} or send a letter to Creative Commons, PO Box 1866, Mountain View, CA 94042, USA. }

\section{Introduction}
\label{sec:introduction}
Several works promote the development of faceted search interfaces \citep{Hearst:06} to reduce information overload and keep users in control of the search process in exploratory search \citep{Marchiorini:06}.
However, most of the existing approaches support an individual user who inspects a single data category, e.g., documents or movies, while pursuing a short-term information goal. 
Similarly, most research on geographic information search focuses on helping individual users retrieve relevant data for particular short-term goals; e.g., finding the available routes between two Points of Interest \citep{Quercia-etal:14}, identifying the 2-star hotels in a specific area \citep{Lionakis-Tzitzikas:17}, or studying the relations among the items of an information category \citep{Andrienko-etal:07}. 

Indeed, map management can go farther than that in order to provide long-term representations of shared projects to users having different information interests. For instance, in participatory decision-making \citep{Coulton-etal:11,Brown-Weber:12}, \cite{Hu-etal:15} point out that 2D maps and 3D virtual environments can facilitate participants' learning and understanding, especially as far as spatial decision-making processes are concerned; see also \citep{Al-Kodmany:99} and \citep{Simpson:01}. 
Moreover, maps can support information sharing and collaboration in simpler and less formal scenarios. For example, if somebody is planning a holiday, a custom map including selected Points of Interest, hotels, and so forth, would provide a personalized projection of the area to be visited that the user can consult and annotate before, during and after the trip, possibly in collaboration with the other people travelling with her/him to gain a common view of the vacation. 

These scenarios suggest the development of custom maps that define Personal Information Spaces \citep{Ardito-etal:13,Ardito-etal:16} useful to organize individual and group activities. For this purpose, maps should be adapted to reflect temporary information goals while persistently storing data in order to facilitate a quick projection and resumption of the collaboration context. 

In this paper we present a faceted exploration model for the management of this type of map. Our model supports a flexible, map-based visualization of heterogeneous data and it enables map focusing to satisfy specific information needs by offering two graphical interactive exploration functions:
\begin{itemize}
    \item 
    The former enables coarse-grained map projection on data categories via opacity tuning, without taking the facets of items into account; all the items of a category are subject to the same visualization policy.
    \item 
    The latter combines opacity tuning with fine-grained faceted search support to enable map projection at different granularity levels, by taking the properties of information items into account.
\end{itemize} 
In both cases, the projection is only visual and the information stored in the map is preserved. The work we present has the following innovative aspects:
\begin{itemize}
    \item 
    {\em Efficient multi-category faceted projection of long-lasting custom maps to answer temporary information needs in sparse and noisy datasets}. Our model suggests information visualization constraints based on attributes of data which support an efficient exploration of the information items stored in the maps. 
    \item 
    {\em Representation of the search context by associating each data category to a compact graphical widget that supports interactive data visualization, faceted exploration, category-based information hiding and transparency of results.} The widgets of the categories searched by the user are located in a side bar of the user interface and play the role of breadcrumbs, representing the types of information that (s)he has explored during the interaction with the system and the applied visualization constraints. 
\end{itemize}
Our model supports geographic information search within the OnToMap collaborative Web GIS \citep{Ardissono-etal:17b,Ardissono-etal:18}.
We tested the model in a user study to assess User Experience and performance in exploratory search. For the experiments we compared different graphical widgets supporting faceted exploration, from traditional ones such as checkboxes, to space-filling ones based on treemaps \citep{Shneiderman:92} and sunburst \citep{Stasko-etal:00} diagrams. 
The study showed that, when working on geographic maps populated with heterogeneous information, our model outperforms simple category-based map projection and traditional faceted search tools such as checkboxes.  Specifically, we obtained the best user performance and experience results using the widget based on the sunburst diagram, which displays visualization criteria in a compact structure.

This article builds on the work described in \citep{Ardissono-etal:18}, which presents our first opacity tuning model. With respect to that work, the present paper introduces graphical widgets that support fine-grained data management and a novel approach to the selection of efficient facets for information exploration in sparse and noisy datasets. The widgets extend the category hiding function provided by the previous model with faceted data exploration to enhance information search and visualization. The present paper also provides an extensive evaluation of the faceted exploration model.

In the following, Section \ref{sec:questions} presents our research questions. Section \ref{sec:related} outlines the related research. Section \ref{sec:OnToMap} overviews OnToMap and Section \ref{sec:model} describes our model. Section \ref{sec:validation} presents the experiments we carried out and Section \ref{sec:discussion} discusses the evaluation results. Section \ref{sec:conclusions} summarizes limitations and future work and the Appendix reports a few tables of details.

\section{Research questions}
\label{sec:questions}
We designed the faceted search model presented in this paper after two preliminary experiments with users carried out in the urban planning domain; see \citep{Voghera-etal:16,Voghera-etal:18}. In those experiments, the projection of long-lasting maps on specific types of information emerged as a useful feature to support data interpretation during project development. This feature was also requested in the final analysis phase, in which human planners analyzed complex maps obtained by integrating the students' projects to identify the most recurring represented territorial elements. 

The present work describes the faceted search support offered by the current version of OnToMap and investigates its usefulness to data search and interpretation in a project map. We pose the following research questions:
\begin{itemize}
    \item[RQ1:] 
    {\em Does faceted exploration of map content help users in finding the needed information in a geographic map that visualizes different types of data?}
    \item[RQ2:] 
    {\em How does a compact, graphical view of the exploration options available to the user, which also shows the status of the information visualization constraints applied to a map, impact on her/his efficiency and experience in data exploration?}
    \item[RQ3:]
    {\em How much does the user's familiarity with the widgets for faceted exploration impact on her/his efficiency in search and on her/his appreciation of the exploration model they offer?}
\end{itemize}
The experiments described in Section \ref{sec:validation} are aimed at answering these questions.

\begin{table*}[h!]
\centering
\caption{Classification of information search models.}
\label{tab:IR-models}
\resizebox{0.45\paperheight}{!}{%
{\def\arraystretch{1.5}
\begin{tabular}{lcccc}

\toprule

Citations & Model & Facets & Filters & 
 \multicolumn{1}{c}{
 \renewcommand{\arraystretch}{0.7}
 \begin{tabular}[c]{@{}c@{}} Visualization \\ of results \end{tabular}}
 \\ 
\midrule  
\citep{Ahlberg-Shneiderman:94} & FilmFinder & 
            \multicolumn{1}{c}{
            \renewcommand{\arraystretch}{0.5}
            \begin{tabular}[c]{@{}c@{}}  static list 
                of terms \end{tabular}} &
            \multicolumn{1}{c}{
            \renewcommand{\arraystretch}{0.5}
            \begin{tabular}[c]{@{}c@{}}  buttons \\ 
                sliders, \dots \end{tabular}} &  
            scatterplot \\
            
\citep{Cao-etal:10} & 
            FacetAtlas & terms &
            - & 
            \multicolumn{1}{c}{
            \renewcommand{\arraystretch}{0.5}
            \begin{tabular}[c]{@{}c@{}}  3D diagram with \\
            multidimensional \\
            relations \end{tabular}} \\  
            
\citep{Cao-etal:11} & 
            SolarMap & terms &
            \multicolumn{1}{c}{
            \renewcommand{\arraystretch}{0.5}
            \begin{tabular}[c]{@{}c@{}}  topic facets + \\
            keyword facets \end{tabular}} & 
            \multicolumn{1}{c}{
            \renewcommand{\arraystretch}{0.5}
            \begin{tabular}[c]{@{}c@{}}  document clusters + \\
            radial facet \\
            visualization \end{tabular}} \\ 

\citep{Chang-etal:19} & 
            SearchLens & terms & lenses & 
            \multicolumn{1}{c}{
            \renewcommand{\arraystretch}{0.5}
            \begin{tabular}[c]{@{}c@{}}  ranked list +\\
            match to 
            lenses\end{tabular}} \\  
           
\citep{Dachselt-Frisch:07} & 
            Mambo &
            \multicolumn{1}{c}{
            \renewcommand{\arraystretch}{0.5}
            \begin{tabular}[c]{@{}c@{}}  from \\
            metadata \end{tabular}} & attributes &
            \multicolumn{1}{c}{
            \renewcommand{\arraystretch}{0.5}
            \begin{tabular}[c]{@{}c@{}} stack-based \\
            hierarchical facets +\\
            zoom on data \end{tabular}} \\    
 
\citep{Dachselt-etal:08} & 
            FacetZoom &
            \multicolumn{1}{c}{
            \renewcommand{\arraystretch}{0.5}
            \begin{tabular}[c]{@{}c@{}}  from \\
            metadata \end{tabular}} & attributes &
            \multicolumn{1}{c}{
            \renewcommand{\arraystretch}{0.5}
            \begin{tabular}[c]{@{}c@{}} stack-based \\
            hierarchical facets +\\
            zoom on data \end{tabular}} \\     
            
\citep{Hearst-etal:02} & Flamenco & 
            \multicolumn{1}{c}{
            \renewcommand{\arraystretch}{0.5}
            \begin{tabular}[c]{@{}c@{}}  
            from  metadata \end{tabular}} &
            \multicolumn{1}{c}{
            \renewcommand{\arraystretch}{0.5}
            \begin{tabular}[c]{@{}c@{}}  text queries \\ 
                hyperlinks \end{tabular}} &  
            \multicolumn{1}{c}{
            \renewcommand{\arraystretch}{0.5}
            \begin{tabular}[c]{@{}c@{}}  matrix view  + miniatures \end{tabular}} \\  
            
\citep{Hildebrand-etal:06} & 
            /facet & 
            \multicolumn{1}{c}{
            \renewcommand{\arraystretch}{0.5}
            \begin{tabular}[c]{@{}c@{}} from RDF \\
            metadata \end{tabular}} &
            \multicolumn{1}{c}{
            \renewcommand{\arraystretch}{0.5}
            \begin{tabular}[c]{@{}c@{}}  attributes + \\
            hyperlinks \end{tabular}} & 
            \multicolumn{1}{c}{
            \renewcommand{\arraystretch}{0.5}
            \begin{tabular}[c]{@{}c@{}}  hierarchical + \\
            facets + 
            list \end{tabular}} \\  

\citep{Hoeber-Dong-Yang:06} & HotMap & terms & text queries & \multicolumn{1}{c}{
            \renewcommand{\arraystretch}{0.5}
            \begin{tabular}[c]{@{}c@{}}  list + term  distribution \\
            with colors \end{tabular}} \\
            
\citep{Hoeber-Dong-Yang:06} & 
            \multicolumn{1}{c}{
            \renewcommand{\arraystretch}{0.5}
            \begin{tabular}[c]{@{}c@{}}  Concept \\ Highlighter \end{tabular}} & concepts & text queries & \multicolumn{1}{c}{
            \renewcommand{\arraystretch}{0.5}
            \begin{tabular}[c]{@{}c@{}}  list + concept  distribution \\
            with colors \end{tabular}} \\
            
\citep{Lee-etal:09} & 
            FacetLens & attributes &
                        \multicolumn{1}{c}{
            \renewcommand{\arraystretch}{0.5}
            \begin{tabular}[c]{@{}c@{}}  text query + \\
            attributes \end{tabular}} & 
            \multicolumn{1}{c}{
            \renewcommand{\arraystretch}{0.5}
            \begin{tabular}[c]{@{}c@{}}  matrix-based \\
            bubbles by filter \end{tabular}} \\ 

\citep{Lionakis-Tzitzikas:17} &
            PFSgeo & 
            \multicolumn{1}{c}{
            \renewcommand{\arraystretch}{0.5}
            \begin{tabular}[c]{@{}c@{}} from RDF\\
            metadata \end{tabular}} &
            \multicolumn{1}{c}{
            \renewcommand{\arraystretch}{0.5}
            \begin{tabular}[c]{@{}c@{}}  attributes + \\
            hyperlinks + \\ geo. dimension \end{tabular}} & 
            ranked list \\      
            
\citep{Papadakos-Tzitzikas:14} &
            Hippalus & 
            \multicolumn{1}{c}{
            \renewcommand{\arraystretch}{0.5}
            \begin{tabular}[c]{@{}c@{}} from RDF \\
            metadata \end{tabular}} &
            \multicolumn{1}{c}{
            \renewcommand{\arraystretch}{0.5}
            \begin{tabular}[c]{@{}c@{}}  attributes + \\
            hyperlinks \end{tabular}} & 
            ranked list \\      
            
\citep{Peltonen-etal:17} & 
            \multicolumn{1}{c}{
            \renewcommand{\arraystretch}{0.5}
            \begin{tabular}[c]{@{}c@{}}  Topic-Relevance \\ Map \end{tabular}} & terms &
            text query & 
            \multicolumn{1}{c}{
            \renewcommand{\arraystretch}{0.5}
            \begin{tabular}[c]{@{}c@{}}  radial distance + \\
            relative distance \end{tabular}} \\

\citep{Petrelli-etal:09} & 
            - & 
            \multicolumn{1}{c}{
            \renewcommand{\arraystretch}{0.5}
            \begin{tabular}[c]{@{}c@{}} from RDF \\
            metadata \end{tabular}} &
            \multicolumn{1}{c}{
            \renewcommand{\arraystretch}{0.5}
            \begin{tabular}[c]{@{}c@{}} attributes + \\
            hyperlinks \end{tabular}} & 
            \multicolumn{1}{c}{
            \renewcommand{\arraystretch}{0.5}
            \begin{tabular}[c]{@{}c@{}}  multiple \\
            visualizations \end{tabular}} \\        
 
\citep{Stadler-etal:14} &
            Facete & 
            \multicolumn{1}{c}{
            \renewcommand{\arraystretch}{0.5}
            \begin{tabular}[c]{@{}c@{}} from RDF\\
            metadata \end{tabular}} &
            \multicolumn{1}{c}{
            \renewcommand{\arraystretch}{0.5}
            \begin{tabular}[c]{@{}c@{}}  attributes + \\
            sem. relations + \\ geo. dimension \end{tabular}} & 
            \multicolumn{1}{c}{
            \renewcommand{\arraystretch}{0.5}
            \begin{tabular}[c]{@{}c@{}}  ranked list + \\
            map-based \\ visualization \end{tabular}}  \\  
 
\citep{Sutcliffe-etal:00} & 
  \multicolumn{1}{c}{
            \renewcommand{\arraystretch}{0.5}
            \begin{tabular}[c]{@{}c@{}}  Thesaurus-\\
            Result Browser \end{tabular}} & terms &
            \multicolumn{1}{c}{
            \renewcommand{\arraystretch}{0.5}
            \begin{tabular}[c]{@{}c@{}}  Hierarchical \\ Thesaurus \end{tabular}} & 
            \multicolumn{1}{c}{
            \renewcommand{\arraystretch}{0.5}
            \begin{tabular}[c]{@{}c@{}}  bullseye \\ browser \end{tabular}} \\
\bottomrule
\end{tabular}%
}
}
\end{table*}

\section{Background and related work}
\label{sec:related}
Exploratory search of large information spaces challenges users in the specification of efficient queries because, as most people are hardly familiar with the search domain, their information goals are often ill-defined  \citep{Marchiorini:06,White-Roth:06}. 
In this paper we focus on faceted search as an alternative, or a complement, to query typing in order to use browsing-based navigation as a proactive guide to information exploration, given the structure of the information space.

Starting from the pioneer filtering model proposed by \cite{Ahlberg-Shneiderman:94}, both \cite{Sacco:00}'s Dynamic Taxonomies and \cite{Hearst:06}'s faceted search model propose to use dynamic filters extracted from items metadata as constraints that the system can suggest to help the user identify relevant terms for information filtering and visualization of results. Specifically, \cite{Hearst-etal:02} present the Flamenco framework in which facet-based filtering is based on the exposure of hierarchical faceted metadata that describes the items of the search domain, i.e., apartments, or images \citep{Yee-etal:03}. 

Researchers also investigate ways to support the specification of the facets to filter results, as well as the access to Semantic Web information and Linked Data \citep{W3C-LinkedData}.
As far as facet specification is concerned, new types of elements are proposed to filter the set of results; e.g., keywords or terms extracted from textual queries, as in HotMap \citep{Hoeber-Dong-Yang:06}, concepts extracted from a document pool, as in Concept Highlighter \citep{Hoeber-Dong-Yang:06}, or terms extracted from a thesaurus as in Thesaurus-Results Browser \citep{Sutcliffe-etal:00}. FacetLens \citep{Lee-etal:09} visualizes clickable facets in matrix-based bubbles, each one associated with a different search filter. Moreover, FacetZoom \citep{Dachselt-etal:08} proposes a stack-based visualization of hierarchical facets, also applied in Mambo \citep{Dachselt-Frisch:07} as a model to combine faceted browsing with zoomable user interfaces. SearchLens \citep{Chang-etal:19} enables users to define long-lasting composite facet specifications (denoted as lenses) to support information filtering on multiple search sessions. In SearchLens, the user can specify the importance of the selected facets; thus, filtering is based on soft constraints used to rank search results.

In the faceted exploration of semantic data \citep{Tzitzikas-Manolis:16}, search interfaces expose rich metadata that support browsing the information space through semantic relations. For instance, in the /facet browser, \cite{Hildebrand-etal:06} propose to combine hierarchical faceted exploration with keyword-based search. Moreover,
\cite{Petrelli-etal:09} enables the user to search for heterogeneous types of information about items (e.g., texts and images) linked according to semantic relations, by extracting facets to guide exploration. Hippalus \citep{Papadakos-Tzitzikas:14} introduces the Faceted and Dynamic Taxonomies to manage both hard and soft constraints in faceted filtering of semantic data and PFSgeo \citep{Lionakis-Tzitzikas:17} extends Hippalus to geographic information management. Finally, focusing on geographic information, \cite{Stadler-etal:14} propose a semantic navigation method for SPARQL-accessible data \citep{SPARQL,GeoSPARQL} in the Facete browser. 

Similar to the cited works, our model exposes metadata derived from semantic knowledge representation. However, it enables users to work on maps populated with multiple data categories, i.e., with heterogeneous information, as well as to focus the maps on temporary interests without losing the overall set of data they contain. This is useful to answer information needs in long-lasting user activities. 
Notice also that OnToMap does not assume to work on RDF data in order to comply with more general data sources, like public crowdmapping platforms, thanks to the mediation of its domain ontology. Moreover, it supports: (i) a browsing-based exploration guided by the structure of the domain ontology, which makes it possible to search for information following both IS-A and semantic relations; (ii) the semantic interpretation of free text queries to identify the data categories (ontology concepts) of interest by abstracting from the specific words occurring in the queries, via Natural Language Processing \citep{Ardissono-etal:16,Mauro-etal:17}.
More generally, OnToMap enables search support over a configurable set of data categories; in this way, it enables complex map development on different information domains. In contrast, most of the previous systems work on a single data type or on a pre-defined set of data categories, as in \citep{Petrelli-etal:09}.  

The dynamic extraction of facets can challenge the user with a large number of browsing options. \cite{Oren-etal:06} focus on the efficiency of exploration and they promote the facets that enable the user to split the set of results in balanced subsets in order to minimize navigation steps. In comparison, we propose a facet selection policy suitable for sparse and highly unbalanced result sets, such as those typically retrieved from crowdsourced data sources, in which very few properties of items split results in subsets having similar cardinality.

Some works propose interactive graphical presentations of keywords to support sensemaking in the exploration of document sets.
For instance, \cite{Peltonen-etal:17} propose the Topic-Relevance Map to summarize on a radial basis the keywords (filters) characterizing the result set, using distance from the center to represent relevance to the search query and angle between keywords to denote their topical similarity. Moreover, FacetAtlas \citep{Cao-etal:10} relates topics in a 3D diagram supporting the representation of multi-dimensional relations among them, and SolarMap \citep{Cao-etal:11} combines topic-based document clustering with a radial representation of facets to support a two-level, topic-based document filtering.
While these models are appropriate to the representation of topics in datasets of unstructured information, they are less relevant to OnToMap, which is fed with structured data and benefits from its domain ontology to organize the presentation of information.

\section{OnToMap overview}
\label{sec:OnToMap}
The OnToMap Web collaborative GIS supports the management of interactive maps for information sharing and participatory decision-making \citep{Ardissono-etal:17b}. A semantic representation of domain knowledge based on an OWL ontology \citep{W3C-OWL} defines the structure of the information space and enables data retrieval from heterogeneous sources by applying ontology mappings; see \citep{Mauro-etal:19b}.
This ontology currently makes it possible to query a dataset of Public Open Data about Piedmont area in Italy and the OpenStreetMap (OSM) server \citep{OpenStreetMap}; in this paper, we focus on OSM data because it is a more general case than the other one. 
The ontology also provides graphical details for map visualization, such as the color and icon associated to each data category; e.g., drugstores are depicted in light green and they are represented as icons marked by a cross.

\begin{figure*}[t]
 \includegraphics[width=1\columnwidth]{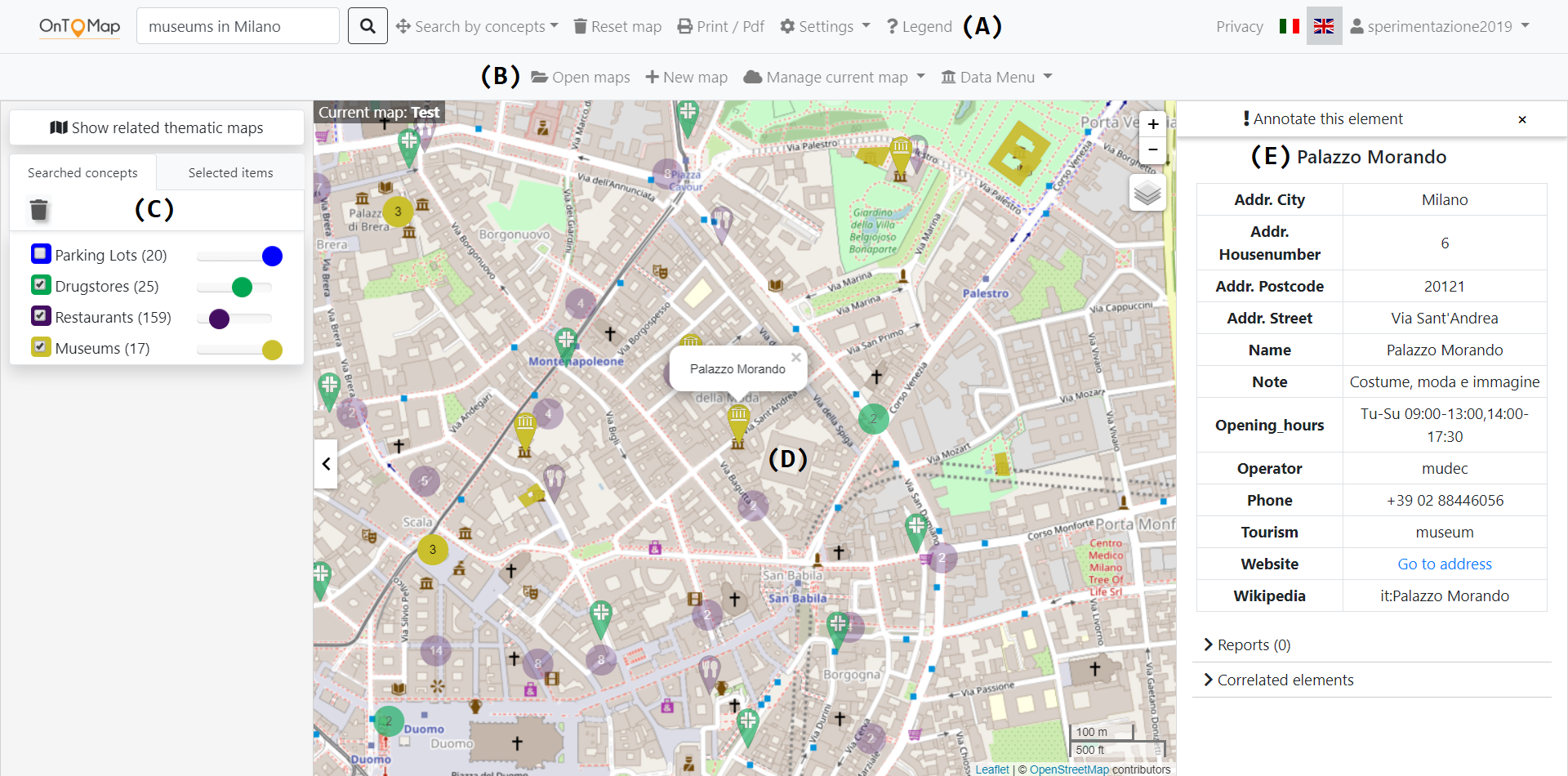}
 \caption{OnToMap user interface showing the widgets based on transparency sliders. 
}
 \label{fig:sliders}
\end{figure*}

OnToMap supports the creation of public and private custom geographic maps to help project design and group collaboration. Search support is based on free text queries and on category browsing.
Textual queries are semantically interpreted using Natural Language Processing techniques with Word Sense Disambiguation \citep{Moro-etal:14}; see \citep{Ardissono-etal:16,Mauro-etal:17}. Data categories can be browsed by means of a simple alphabetical menu with auto-completion or by navigating a graphical representation of the taxonomy defined in the domain ontology.

Figure \ref{fig:sliders} shows the user interface of OnToMap. 
The top bar includes the control panel that supports (A) free text search (``Search...'') and category browsing (``Browse by concepts''), basic map management and user authentication; (B) map management tools available to authenticated users. The left side bar (C) displays the widgets of the data categories that the user has searched for during the interaction: a different widget is associated to each category in order to regulate the visualization of its items in the map. 

The main portion of the interface (D) contains the geographic map, which displays information items as pointers with category-specific icons or as geometries, depending on the input data. Color coding \citep{Hoeber-Dong-Yang:06} visually connects the widgets representing data categories in the side bar with the corresponding items in the map. Moreover, when several items of a category are located in a restricted area, a cluster colored as the category is displayed to avoid cluttering and, at the same time, provide visual information about the grouped items.
If the user clicks on the visual representation of an item, the system generates a table (E) describing its details.

\section{Information exploration model}
\label{sec:model}

Our information exploration model is integrated in the OnToMap system to support information search and it includes two main types of functions, implemented as interactive graphical widgets.

\subsection{Exploration function 1: coarse-grained map projection by means of transparency sliders}
\label{sec:transparencySliders}
This function, introduced in \citep{Ardissono-etal:18}, supports map projection via opacity tuning: for each searched category, a transparency slider enables the user to assign different levels of opacity to its items; the widget also has a checkbox to temporarily hide information by means of a click, without changing the degree of opacity selected for the category. 

The side bar of Figure \ref{fig:sliders} shows the widgets based on transparency sliders. In the map, museums are visualized in full color because the slider of the ``Museums'' category is selected and tuned to maximum opacity. Differently, drugstores and restaurants are semi-transparent and the map hides the items of the ``Parking Lots'' category because its slider is de-selected. 

The transparency slider does not enable the specification of constraints on facet values; i.e., it works at the granularity level of the represented category and it uniformly tunes the opacity of items. Nevertheless, this widget supports visual simplification by enabling the user to temporarily hide information by type. Basically, opacity tuning enables her/him to highlight the information in focus while maintaining an overview of what has been searched in the map.
This model is inspired by \cite{Colby-Sholl:91}'s work on layers visualization but it separately handles the opacity of items belonging to different categories; moreover, it supports the visualization of multiple layers, as a generalization of Translucent Overlay \citep{Lobo-etal:15}.

\subsection{Exploration function 2: faceted approach}
\label{sec:facetedExploration}
\begin{figure*}[t]
 \includegraphics[width=1\columnwidth]{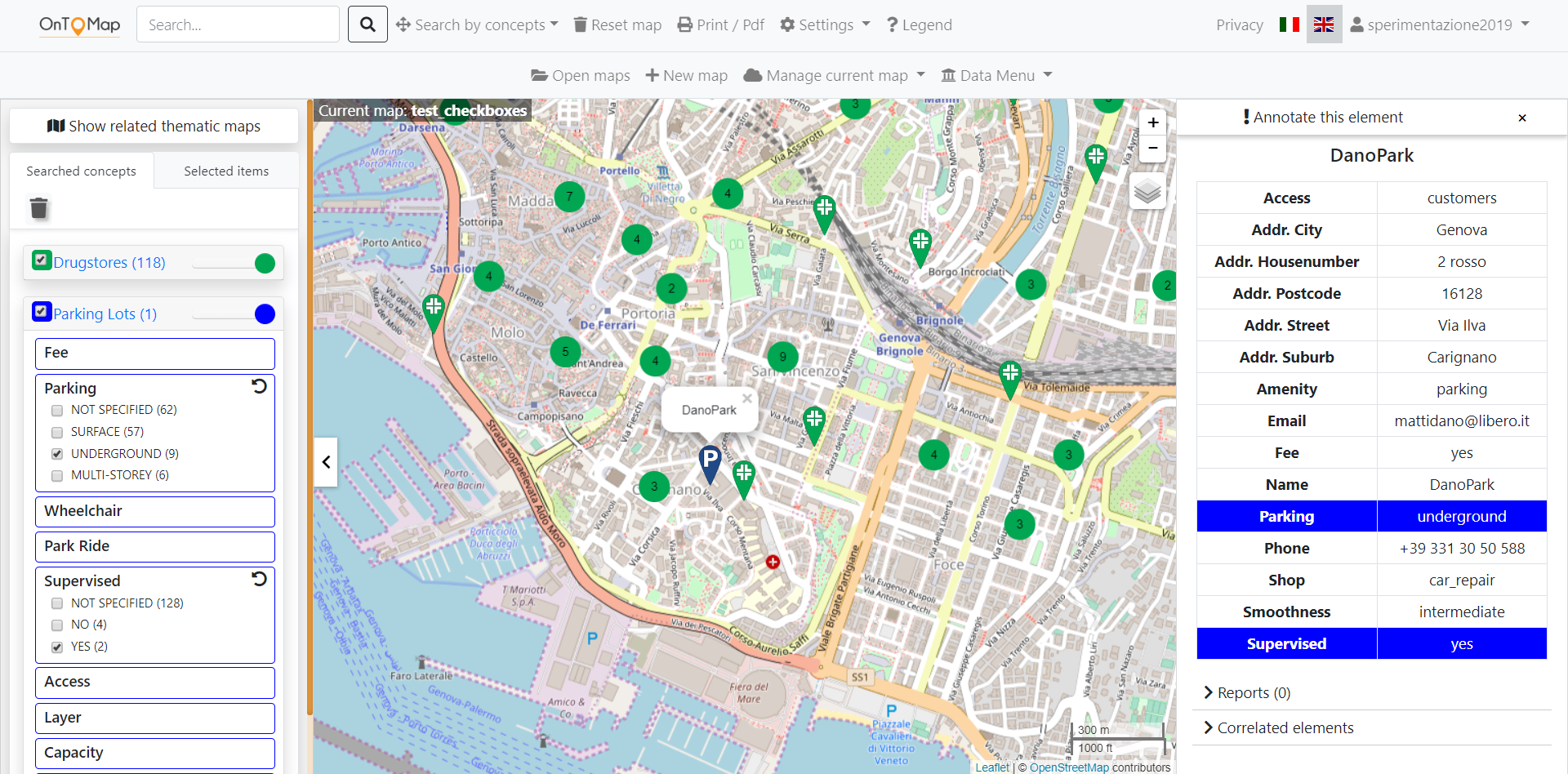}
 \caption{User interface showing the widgets based on checkboxes.}
 \label{fig:checkboxes}
\end{figure*}

This function combines coarse-grained and fine-grained specification of visualization constraints by integrating transparency sliders with faceted information exploration.
The widgets implementing this function include a transparency slider and an internal component showing the facets of the represented category. The internal component can be a set of checkboxes, a treemap or a sunburst diagram, depending on the layout selected for the user interface, and it enables the user to specify visualization constraints based on facet values. The transparency slider works in combination with facet selection and tunes the opacity of the visualized items. 
The widgets are interactive and they can be opened or closed by clicking on them; a closed widget only shows its own transparency slider; e.g., see ``Drugstores'' in Figure \ref{fig:checkboxes}, which shows the layout based on checkboxes.

Let us consider a facet $f$ of a category $C$ and the set of retrieved items that belong to $C$, henceforth denoted as $E_C$, i.e., extension of $C$. The visualization of the values of $f$ in the widget is aimed at providing the user with a preview of the corresponding items in the map. For this purpose we adopt a standard approach to facet suggestion \citep{Oren-etal:06,Hearst:06}: 
\begin{itemize} 
  \item 
  The widget only displays the values $\{v_{1}, \dots, v_{m}\}$ of $f$ that have at least one item in $E_C$ to prevent the user from following links to zero solutions. 
  \item 
  The values of $f$ are sorted from the most frequent to the least frequent ones in $E_C$. Moreover the widget shows, or makes available on mouse over, the number of items corresponding to each value. Notice that the widget may also show a ``NOT SPECIFIED'' value to represent the subset of items in which $f$ is not defined. This is aimed at providing the user with a visual representation of the coverage of the facet in the results.
  \item 
  In order to limit visual complexity, long lists of values are dropped, making their tails available on demand by providing a ``More...'' link or a ``+'' symbol, depending on the layout of the widget.
  \end{itemize}
By default, none of the facets in the widget of a category $C$ is selected.
If the user picks one or more values of the same facet, this is interpreted as an OR constraint because (s)he has specified that all those values are eligible for visualization. 
Conversely, the selection of values that belong to distinct facets of $C$ generates an AND constraint because it identifies the items having more than one property restricted to specific values. 
For instance, if the user chooses $f_i = v_{i1}$, $f_i = v_{i2}$ and $f_j = v_{j1}$, items $\{x \in E_C~|~ f_i(x) \in \{v_{i1}, v_{i2}\} \wedge f_j(x) = v_{j1}\}$ are shown and the other items are hidden. 

We use color coding to link visualization constraints to map content: the tables showing the details of items highlight the facets corresponding to the selected visualization constraints in the color associated to the category. In this way, the user can quickly identify the characteristics that make items eligible for being displayed. For instance, the table of ``DanoPark'' in Figure \ref{fig:checkboxes} has the ``Parking'' and ``Supervised'' facets highlighted in blue because they correspond to the visualization constraints imposed on the ``Parking Lots'' widget.

\begin{figure*}[t]
 \includegraphics[width=1\columnwidth]{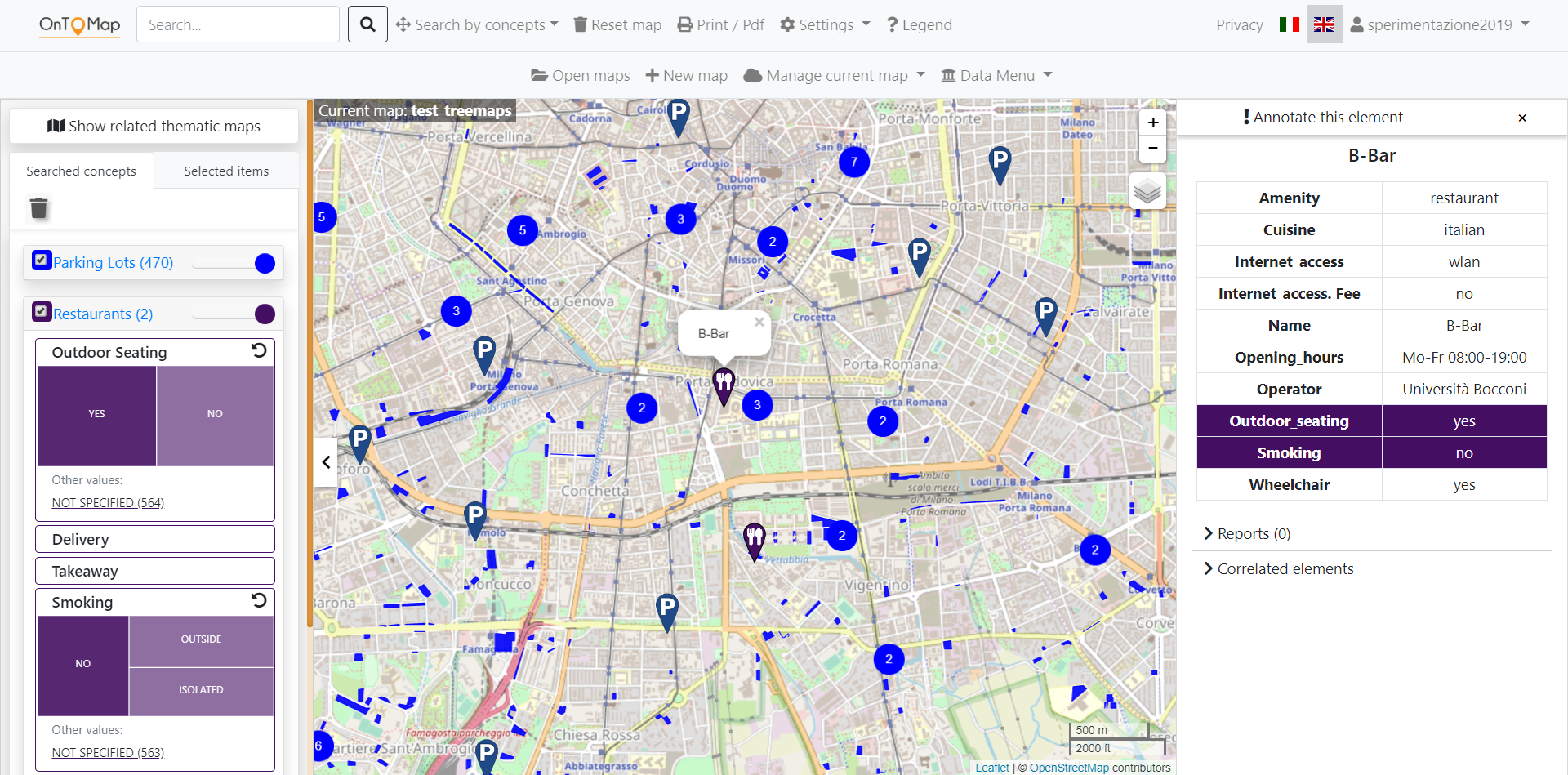}
 \caption{User interface showing the treemaps as faceted exploration widgets.}
 \label{fig:treemaps}
\end{figure*}

\begin{figure*}[t]
 \includegraphics[width=1\columnwidth]{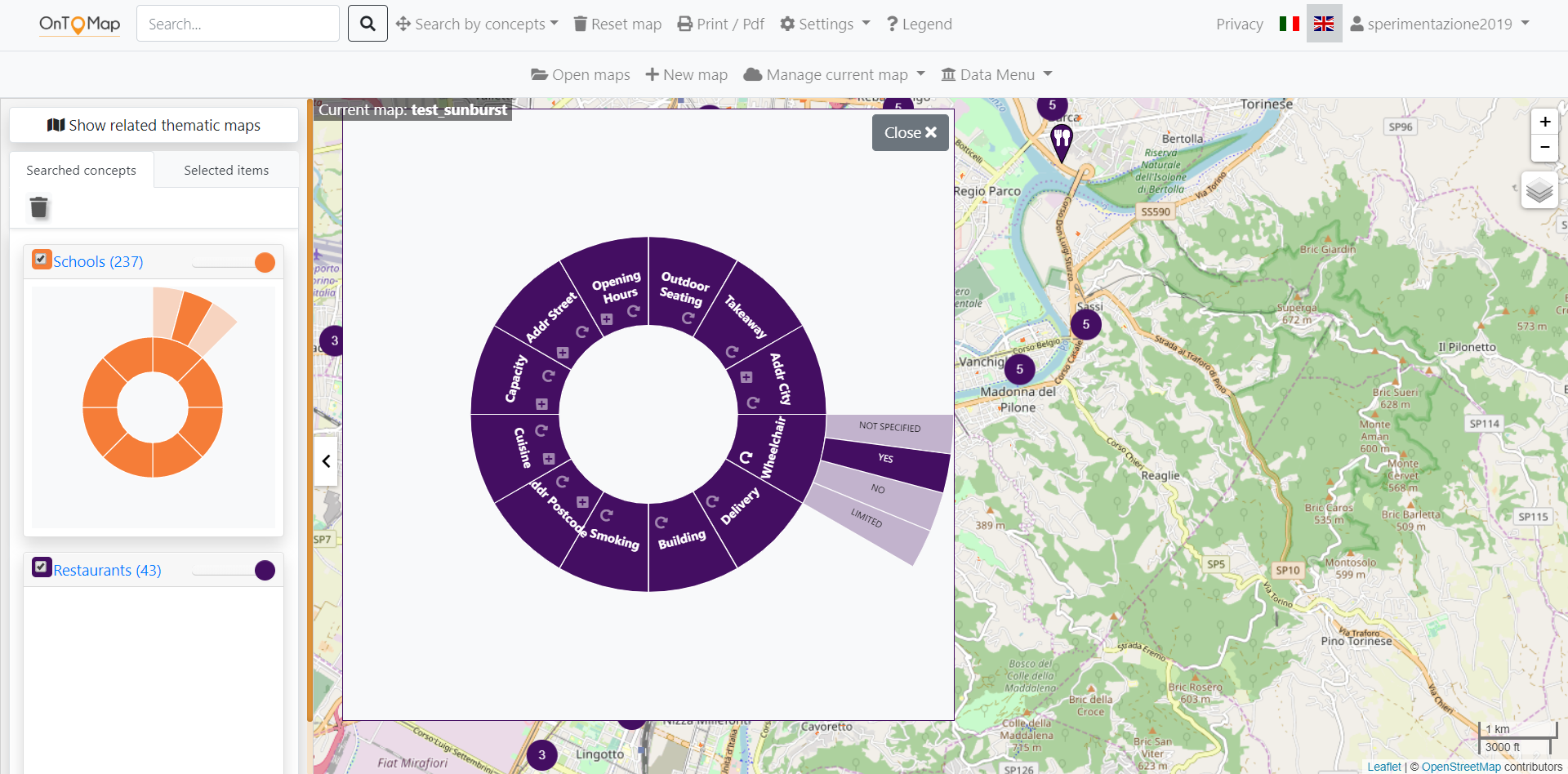}
 \caption{User interface showing the sunburst as faceted exploration widgets.}
 \label{fig:sunburst}
\end{figure*}

\subsubsection{Layouts of the widgets for faceted information exploration}
\label{sec:widgets}
Before providing details about how we select the facets to be included in the widgets we present the layouts we developed. 
\begin{itemize}
    \item 
    The {\bf widgets based on checkboxes} contain a rimmed rectangle for each facet to be shown. By clicking on a rectangle, the user expands (or closes) the corresponding facet. An expanded facet shows values as checkboxes and offers a ``More...'' link to show the hidden ones. 
    For instance, Figure \ref{fig:checkboxes} shows eight facets of widget ``Parking Lots": ``Fee'', \dots, ``Capacity''. 
    Users can select the checkboxes to impose visualization constraints in the map. In the figure the user has expanded the ``Parking'' and ``Supervised'' facets and (s)he has selected values ``UNDERGROUND'' and ``YES''.
    \item
    {\bf The widgets based on treemaps} include facets in rimmed rectangles as well. However, when a facet is expanded, its values are displayed as components of a treemap whose size depends on the cardinality of the corresponding set of items (larger size means larger cardinality); see Figure \ref{fig:treemaps}. Long values are shortened but they can be visualized, together with the cardinality of the corresponding sets of items, on mouse over. Only the most frequent values are included in the treemap; the other ones are available below it or on demand (``More...'') and the user can add them to the treemap by means of a click.
    The user can (de)select values by clicking on them. The selected values take the color of the category (e.g., ``YES'' in ``Outdoor Seating'' and ``NO'' in ``Smoking''); the other ones have a pale tone of the same color.
   \item
    The {\bf widgets based on the sunburst diagram} show the facets of the represented category $C$ in a ring having the color associated to $C$. The diagram is visualized in a pop-up window that the user can open or close by means of a click, and the side bar of the user interface only displays the thumbnails of the sunbursts; see Figure \ref{fig:sunburst}. 
    The user can expand each facet by clicking on the portion of ring representing it: values are shown in a second level, sorted clockwise by decreasing frequency in the extension of $C$. Only the most frequent values are shown but the user can view and add the other ones by clicking on the ``+" button located in each portion of the internal ring; the sunburst is extended by adapting the size of the displayed values. The user can (de)select values by clicking on them and color coding is applied to link visualization constraints to map content. 
\end{itemize}

\subsection{Selection of facets to be included in the information exploration widgets}
The dynamic generation of widgets for the exploration of search results retrieved from open data sources is challenged by the amount and variability of the information items to be managed. Facets have thus to be analyzed in order to identify the most convenient ones for map content analysis. 

\subsubsection{Navigation quality in semantic data repositories}
\cite{Oren-etal:06} introduce the {\em navigation quality} of a facet $f$ to describe its efficiency in supporting information browsing of RDF data repositories. This measure takes values in [0, 1] (where 1 is the best value) and is based on the product of three metrics, which take values in [0, 1] as well:
\begin{enumerate} 
\item 
The {\em balance} of $f$, i.e., its capability to split results in subsets having similar cardinality; equally distributed facets have maximum balance.
\item 
An inverse measure of the {\em number of distinct values} of $f$ occurring in the results, denoted as {\em ``object cardinality''}. The authors consider as acceptable the facets that have between 2 and 20 values because they can be displayed in a search interface without overloading the user. 
\item 
The {\em frequency} of $f$ in the results, i.e., the percentage of retrieved items in which the value of $f$ is specified. 
\end{enumerate}
\begin{table*}[!t]
\centering
\caption[Caption]{Value distribution of facets ``Cuisine'', ``Outdoor Seating'' and ``Takeaway'' (OSM tag: ``amenity=restaurant'') in Torino city bounding box; retrieved using Overpass Turbo \citep{OverpassTurbo} on Sept. 20th, 2019. The results include 719 items, out of which 432 specify the value of ``Cuisine'', 92 specify the value of ``Outdoor Seating'' and 76 specify the value of ``Takeaway''.
}
\label{tab:cuisine}
\subfloat{
\resizebox{0.41\textwidth}{!}{%
{\def\arraystretch{0.7}
\begin{tabular}{lc}
\toprule
Cuisine  & Count
\\ 
\midrule  
PIZZA  & 111 
\\
ITALIAN  & 74           \\
REGIONAL & 37           \\
CHINESE  & 28          \\
JAPANESE & 17         \\
ITALIAN; PIZZA & 15    \\
SUSHI  & 11           \\
ITALIAN; REGIONAL & 10         \\
PIZZA; ITALIAN & 10             \\
MEXICAN & 9          \\
KEBAB & 7           \\
INDIAN  & 4          \\
ASIAN  & 3            \\
CHINESE; JAPANESE   & 3            \\
FISH & 3       \\
INTERNATIONAL & 3          \\
ITALIAN; PIZZA; REGIONAL  & 3       \\
ITALIAN\_PIZZA  & 3            \\
PERUVIAN & 3           \\
STEAK\_HOUSE  & 3             \\
AMERICAN  & 2          \\
CHINESE; PIZZA  & 2           \\
GREEK  & 2           \\
ITALIAN\_PIZZA; PIZZA  & 2      \\
KEBAB; PIZZA  & 2              \\
LOCAL & 2            \\
MEDITERRANEAN  & 2             \\
PIZZA; KEBAB  & 2              \\
REGIONAL; ITALIAN  & 2       \\
AFRICAN  & 1           \\
...  &           \\
\multicolumn{2}{l}{\textit{56 more values with Count=1.}}   \\

\bottomrule
\end{tabular}%
}
}
}
\subfloat{
\resizebox{0.27\textwidth}{!}{%
{\def\arraystretch{0.7}
\begin{tabular}{lc}
\toprule
Outdoor Seating & Count 
\\ 
\midrule  
NO  & 59 
\\
YES  & 33     \\
\bottomrule
\end{tabular}%
}
}
}
\subfloat{
\resizebox{0.21\textwidth}{!}{%
{\def\arraystretch{0.7}
\begin{tabular}{lc}
\toprule
Takeaway & Count 
\\ 
\midrule  
YES      & 62         
\\
NO       & 10           \\
ONLY     & 4     \\
\bottomrule
\end{tabular}%
}
}
}
\end{table*}
Navigation quality cannot be applied in OnToMap because of its assumptions: firstly, statistics about OSM data provided by TagInfo \citep{OSMtaginfo} show that most of the tags are hardly used.\footnote{For instance, by invoking \url{https://taginfo.openstreetmap.org/tags/amenity=restaurant\#combinations} it is possible to learn that ``amenity=restaurant" has 178 different tags, only 36 of which occur in more than 2\% of the items mapped in OSM worldwide. Moreover, the most frequent tag is ``name'', which is only defined in 90.92\% of items, in spite of its importance as a POI identifier.} This can be explained because crowdmappers tend to underspecify the items they map; moreover, they sometimes define new tags instead of using the existing ones, thus generating a plethora of synonyms which increase in an uncontrolled way the number of distinct facets and values. This phenomenon is so widespread that several efforts try to systematize OpenStreetMap through semantic knowledge representation; e.g., see \citep{Codescu-etal:11,Ballatore-etal:13}.
We also notice that several results retrieved from OSM are unbalanced and can be split into (i) a large set of items in which the facet is not available, (ii) a few values identifying sets of items with reasonable cardinality, and (iii) a long tail of values represented by one or two items. For instance, Table \ref{tab:cuisine} shows the distribution of three facets retrieved from OSM by searching for ``amenity=restaurant'' (which corresponds to the ``Restaurants'' category of the OnToMap domain ontology) on Torino city bounding box. The facets have fairly poor coverage and they are unbalanced: ``Cuisine'' is specified in 432 items out of 719 and it exhibits a distribution with a long tail; ``Outdoor Seating'' and ``Takeaway'' are specified in 92 and 76 items respectively; ``Name'', not shown, is balanced but it only occurs in 675 items.

We thus define a novel approach to the computation of facet efficiency that suits these types of distribution and is robust towards information lack. The idea is that (i) coverage has to be taken into account as a separate factor to select useful facets, and (ii) balance and number of values have to be controlled by the cardinality of the sets of items identified by the facet.

\subsubsection{Our approach: evaluating exploration cost in sparse, unbalanced datasets}
When searching for information in crowdsourced data sources, the suggestion of the most representative facet values in a result set is a primary goal because it enables the system to provide the user with a relevant number of items to choose from. Moreover, it can be complemented by free text queries that let the user express specific information needs; e.g., in OnToMap free text queries support the retrieval of very specific items, such as ``{\em Pediatric} hospitals in Torino''. It thus makes sense to propose facets that, regardless of balance, identify some fairly large subsets of items, possibly leaving the long tail apart or making it available on demand; e.g., consider the first values of ``Cuisine'' in Table \ref{tab:cuisine}. Given these premises, we propose a two-step evaluation of facets efficiency to exploration support.

1) In the first step, we consider {\em frequency} as a pre-filtering metric to exclude from any further computation the facets that appear very rarely in the results.
Having sampled a set of queries to OSM and taking Taginfo statistics as a baseline, we empirically set to 3\% the minimal frequency threshold under which a facet is considered as useless. Only the facets over this threshold are considered for the evaluation of their efficiency.

2) In the second step, given the highly variable distribution of facets, we consider balance and number of values in combination.
We are interested in facets that split $E_C$ in at least some portions having significant cardinality because they identify homogeneous, relatively large sets of items to be analyzed. These facets enable the system to propose visualization criteria that significantly reduce the search space by showing the most representative values, leaving the other ones on demand. Differently, in a small result set, as those typically retrieved when the selected bounding box is strict, there are few items; therefore, the efficiency in splitting results is less important because the user can easily analyze items one by one. 
In order to capture this intuition, we compute the {\em cost of exploring the extension} $E_C$ of a category $C$ by means of a facet $f$ that takes values in $\{v_1, \dots, v_m\}$ as follows:
   \begin{equation}
        explorationCost(f) = \frac{-\sum\limits_{j=1}^{m} p(v_{j}) log_2 p(v_{j})}{meanCard(f)}
        \label{eq:cost}
    \end{equation}
$explorationCost(f)$ takes values in ${\rm I\!R}^{+}$. In the formula, $p(v_{j})$ is the probability of $v_{j}$ in $E_C$, computed by considering the values $v_{j} \neq$ ``NOT SPECIFIED''. Moreover, $meanCard(f)$ is the mean frequency of the values of $f$ in $E_C$, i.e., the mean cardinality of the subsets of results identified by $f$: 
    \begin{equation}
        meanCard(f) = \frac{|E_C|}{m}
    \end{equation}
The components of Equation \ref{eq:cost} have the following roles:
\begin{itemize}
    \item 
    The numerator represents the (not normalized) entropy of $f$, which takes values in ${\rm I\!R}^{+}$. The entropy of an information source is an average measure of the amount of uncertainty of its own $m$ symbols; it is positively influenced by both the number of values that  the source can take and by the balance of the corresponding subsets of items. For instance, given two balanced facets $f_1$ and $f_2$, if $f_2$ has more values than $f_1$, $f_2$ also has higher entropy than $f_1$. Moreover, if two facets have the same number of values, the most balanced one has the highest entropy. Finally, if all the items of $E_C$ have the same value of $f$ (e.g., all the schools located in the bounding box are primary ones), the entropy is $0$, meaning that the facet does not help discriminate among the items of $E_C$.
    \item 
    The denominator of Equation \ref{eq:cost} captures our interest in the facets that split results in fairly large subsets: even though a facet $f$ has high entropy (e.g., because it has several values), its cost is smoothed if the subsets of items it identifies have high mean cardinality, because $f$ enables the user to browse a large portion of results in few steps.
\end{itemize}

\definecolor{11}{HTML}{60C079}
\definecolor{12}{HTML}{F2EA7F}

\definecolor{21}{HTML}{6BC278}
\definecolor{22}{HTML}{FBC57D}

\definecolor{31}{HTML}{BFD67E}
\definecolor{32}{HTML}{F6E58B}

\definecolor{41}{HTML}{FFE283}
\definecolor{42}{HTML}{F86A69}

\definecolor{51}{HTML}{FFD380}
\definecolor{52}{HTML}{F69574}

\definecolor{61}{HTML}{F66A69}
\definecolor{62}{HTML}{F66A69}

\begin{table*}[t]
\small
\centering
\begin{tabular}{lcc}
\toprule
  & Exploration cost &  \begin{tabular}[c]{@{}c@{}}(1 - navigation quality) \end{tabular} \\ \midrule  

Outdoor Seating & \cellcolor{11}0.0205           & \cellcolor{12}0.8924            \\
Takeaway        & \cellcolor{21}0.0335         & \cellcolor{22}0.9497           \\
Ex1             & \cellcolor{31}0.1500      & \cellcolor{32}0.8949           \\
Cuisine         & \cellcolor{41}0.8738           & \cellcolor{42}1.0000                  \\
Ex2             & \cellcolor{51}1.0000         & \cellcolor{52}0.9842           \\
Name            & \cellcolor{61}9.3987         & \cellcolor{62}1.0000       \\
\bottomrule
\end{tabular}
\captionof{figure}{Exploration cost and complement of navigation quality of a set of facets. The color scale varies from the lowest cost values, depicted in green, to the highest ones, in red. Notice that colors are tuned to the values observed in this example; i.e., [0, 10] for exploration cost and [0, 1] for the complement of navigation quality.}
\label{fig:comparison}
\end{table*}

\noindent
Figure \ref{fig:comparison} graphically compares the exploration cost of Equation \ref{eq:cost} with \cite{Oren-etal:06}'s navigation quality on a few facets; see Tables \ref{tab:costdetails} and \ref{tab:navqdetails} in the Appendix for details. We consider ``Cuisine'', ``Takeaway'', ``Outdoor Seating'' and ``Name'', based on the data described in Table \ref{tab:cuisine}, and two toy examples:
\begin{itemize}
    \item 
    ``Ex1'', specified in 160 items, has 8 distinct balanced values, with $meanCard=20$.
    \item
   ``Ex2'', specified in 24 items, has 8 distinct balanced values, with $meanCard=3$.
\end{itemize}
Notice that Oren and colleagues compute a quality measure, i.e., the highest values are the preferred ones; conversely, we compute a cost function that has the opposite interpretation. In order to facilitate the comparison, Figure \ref{fig:comparison} graphically shows the {\em complement of navigation quality} in the [0, 1] interval and it tunes the color scale to the values observed in this example; i.e., [0, 10] for exploration cost and [0, 1] for the complement of navigation quality.\footnote{\cite{Oren-etal:06}'s model introduces the $\sigma$ and $\mu$ parameters for the computation of balance and object cardinality metrics but we could not find the exact values that they applied in their experiments. We reproduced the expected behavior, following the indications given in the paper, by setting $\mu=2$ and $\sigma=4.9$.} 
\begin{itemize} 
\item 
In both approaches ``Name'' has very high cost, which is desirable because this facet identifies hundreds of subsets of items to be browsed one by one. 
\item 
According to \citep{Oren-etal:06}, ``Outdoor Seating'', ``Ex1'' and ``Takeaway'' are moderately inefficient, and ``Takeaway'' has higher cost than the other ones; the reason is the low coverage of these facets and, with the exception of ``Ex1'', their lack of balance. Differently, our model attributes low cost to these facets because they have few values which represent non elementary sets of solutions to be inspected.
\item 
The main disagreement is in the evaluation of ``Cuisine'' and ``Ex2''. According to \citep{Oren-etal:06}, ``Cuisine'' is totally inefficient because of its partial coverage of items, lack of balance and high number of values. Moreover, ``Ex2'' is penalized by the lack of coverage of results. In our approach ``Cuisine'' has moderate cost, in spite of the many values it can take, because it identifies a few large subsets that deserve attention when browsing results, and the long tail of the facet can be ignored. ``Ex2'' has higher cost than ``Cuisine'' because it identifies very small sets of items.
\end{itemize}
In summary, our approach supports the identification of facets which are not ``perfect'' from the {\em divide et impera} viewpoint because they only occur in a subset of results and/or they split data in an unbalanced way. However, it works on realistic cases in which balanced, frequent facets are extremely rare. Moreover, it promotes facets that split results in subsets having a significant cardinality because they are valuable for browsing results.

\subsection{Selection of facets to be included in the widgets}
\label{sec:facetSelection}
In order to select the facets to be shown in the widgets, we first exclude those having $cost(f)=0$ because this means that they have a single value in $E_C$. Then, we sort facets by increasing cost and we include them in the widget up to a maximum number of 12 to avoid cluttering the user interface. 

By applying Equation \ref{eq:cost} to the results of query ``amenity=restaurant'' on Torino city bounding box, we obtain the following sorted list of facets: 
 ``Outdoor Seating'', ``Takeaway'', ``Wheelchair'', ``Delivery'', ``Addr city'', ``Smoking'', ``Building'', ``Addr postcode'', ``Cuisine'', ``Capacity'', ``Addr street'' and ``Opening hours''; see Figure \ref{fig:sunburst}. Almost all these facets correspond to semantically relevant dimensions. Only ``Addr city'' seems useless because the query is bounded in Torino city; however, according to the geocoder we use, the area of the map that is considered includes Torino and a few small cities in its boundary. Other facets, such as ``Name'', ``Phone'' and ``URL'', are excluded from the sunburst because they have very high cost (they are identifiers) and thus take the final positions in the ranked list. Facets such as ``Cuisine 1'', which is redundant with respect to ``Cuisine'', are excluded because they are below the minimum coverage threshold. Indeed, ``Cuisine 1'' is the typical tag that somebody has duplicated instead of using the main ``Cuisine'' one.

\section{Validation of our faceted exploration model}
\label{sec:validation}
\subsection{Study design}
We conducted a user study by exploiting OnToMap to evaluate the four types of information exploration widgets described in Section \ref{sec:model}, as far as data interpretation in a geographic map is concerned. Specifically, we were interested in comparing:
\begin{itemize}
    \item 
    The exploration model based on transparency sliders (which supports information hiding at the granularity level of the data category) to the more expressive one that also supports faceted exploration.
    \item 
    The alternative graphical models we defined for faceted information exploration in order to understand which ones are more effective to help users in the exploration of an information space via map projection.
\end{itemize}
For the experiment we defined a simple project planning scenario concerning the preparation of a tourist trip in Torino city. We instructed each participant that (s)he should imagine to plan a tour with some friends in different areas of the town. We also explained that, for each area, (s)he would find the information about Points of Interest that might be visited (e.g., urban parks, monuments, etc.), as well as travel facilities (e.g., parking lots), by exploring a custom geographic map focused on the specific area and previously prepared by her/his friends. We aimed at separately evaluating the four widget types we defined but we wanted to minimize the learning effect on participants. Therefore, we prepared four maps, each one focused on a different geographic area of Torino city. Each map was populated with multiple data categories representing Points of Interest and travel facilities.
We investigated participants' performance and User Experience in four map learning tasks, each one using a different type of widget and map:
\begin{itemize}
    \item {\em Task1}: question answering using checkboxes in combination with transparency sliders.
    \item {\em Task2}: question answering using treemaps in combination with transparency sliders.
    \item {\em Task3}: question answering using sunburst in combination with transparency sliders.
    \item {\em Task4}: question answering using transparency sliders.
\end{itemize}
The study was a within-subjects design one. 
We considered each treatment condition as an independent variable and every participant received the 4 treatments. We counterbalanced the order of tasks to minimize the impact of result biases and the effects of practice and fatigue.
People participated in the user study on a voluntary basis, without any compensation, and they signed a consent to the treatment of personal data. 
The participation to the user study took place live, i.e., we did not perform any online interviews.

\subsection{The experiment}
One person at a time performed the study which lasted about 30 minutes.
Before starting the user study, the participant watched a video describing the widgets and showing how they work. After that, (s)he interacted with OnToMap on an sample map to get acquainted with the user interface of the system. We did not impose any restrictions on this activity and we allowed the participant to take as much time as (s)he needed in order to comply with diverse backgrounds and levels of confidence with technology.
Then, we asked her/him to answer a pre-test questionnaire designed to assess demographic information, cultural background, as well as familiarity with map-based online applications.

During the study, we asked the participant to use OnToMap in the context of the organization of the trip. For each task (s)he had to look at the associated map and (s)he had to answer two questions which required counting elements that have certain properties, or identifying items given their descriptions. For each specific map, all participants answered the same two questions.
As far as counting is concerned, we forced the participant to analyze the map by asking her/him to answer the questions in a geographic area delimited by an orange border. In this way, (s)he could not simply read the cardinality information provided by the faceted exploration widgets, which work by taking the bounding box of the map as a reference to specify how many items satisfy the selected visualization constraints.
The questions proposed to the participants had the following templates:
\begin{itemize}
\item
How many \textit{category name} having $characteristic_1$ and/or \dots and/or
\linebreak 
$characteristic_n$  are visualized within the area delimited by the orange line in the map? 

For instance, ``How many Christian churches accessible to wheelchairs are visualized within the area delimited by the orange line in the map?''. In the question, ``Christian'' is a value of facet ``Religion'' and wheelchair accessibility corresponds to value ``YES'' of facet ``Wheelchair''.
\item
Find \textit{category name} having
$characteristic_1$ and/or \dots and/or
\linebreak
$characteristic_n$ within the orange line in the map, and list them.

E.g., find restaurants serving pizza or Italian food (values of ``Cuisine'').
\end{itemize}
In {\em Task1}, {\em Task2} and {\em Task3}, we proposed selective questions because we wanted to understand whether the widgets helped participants satisfy specific information needs by exploring the metadata of the searched categories and by projecting the maps accordingly.
Differently, the questions of {\em Task4} did not require the imposition of any visualization constraints because participants only used the transparency sliders; in this task we assessed the general usefulness of category-based map projection in reducing the visual complexity of a map that includes diverse types of information. This function was appreciated by users in a previous experiment \citep{Ardissono-etal:18} but we wanted to evaluate it extensively.

While the participant carried out a task, the experimenter took notes about how much time (s)he used to answer the questions, sitting at some distance from her/him. We did not put any time restrictions on question answering and we allowed checking the answers multiple times.

\begin{table*}
\centering
\caption{Post-task questionnaire (translated from the Italian language).}
\resizebox{\columnwidth}{!}{%
{\def\arraystretch{1.3}
\begin{tabular}{ll}
\toprule
\# & Question \\ \midrule
1 & 
\multicolumn{1}{l}{
\renewcommand{\arraystretch}{0.7}
\begin{tabular}[l]{@{}l@{}}How familiar are you with the widget that you just used? \end{tabular}}
  \\ 
2 &\multicolumn{1}{l}{
\renewcommand{\arraystretch}{0.7}
\begin{tabular}[l]{@{}l@{}}How much did the widget help you find the information that you were looking for in the\\ map?\end{tabular}}
  \\
3 & 
\multicolumn{1}{l}{
\renewcommand{\arraystretch}{0.7}
\begin{tabular}[l]{@{}l@{}}How much did the widget help you save effort in answering the questions we asked you? \end{tabular}}
  \\ 
4 & 
\multicolumn{1}{l}{
\renewcommand{\arraystretch}{0.7}
\begin{tabular}[l]{@{}l@{}}Please, rate the ease of use of the widget you just used. \end{tabular}}
   \\ 
5 &  
\multicolumn{1}{l}{
\renewcommand{\arraystretch}{0.7}
\begin{tabular}[l]{@{}l@{}}Please, rate the novelty of the widget you just used.  \end{tabular}}
   \\ 
6 & 
\multicolumn{1}{l}{
\renewcommand{\arraystretch}{0.7}
\begin{tabular}[l]{@{}l@{}}Did you encounter any difficulties in finding the information that you were \\ looking for? \end{tabular}}
 \\
7 & 
\multicolumn{1}{l}{
\renewcommand{\arraystretch}{0.7}
\begin{tabular}[l]{@{}l@{}}Is there any aspect of the widget you used that you particularly appreciated? \end{tabular}}
 \\ \bottomrule
\end{tabular}
}}
\label{tab:questions}
\end{table*}

As objective performance indicators, we measured task completion time and the percentage of correctly answered questions. As a subjective measure, we analyzed User Experience: after the completion of each task, the participant filled in a post-task questionnaire to evaluate the type of widget (s)he had just used. We were interested in evaluating the following traits of the facet-based widgets: familiarity, helpfulness in finding information, effort saving in solving the task to be performed, ease of use and novelty.  
We defined the questions to be posed by taking inspiration from NASA TLX questionnaire \citep{NASA-TLX}; however, for simplicity, we kept a 5-points Likert scale for the expression of ratings. Table \ref{tab:questions} shows our questionnaire:
for questions 1-5 the participant had to provide values from 1, the worst value, to 5, the best one; questions 6 and 7 were open to free text comments.

After the completion of the four tasks the participant filled in a post-test questionnaire to compare the widgets.\footnote{Also in this case, we took inspiration from User Experience Questionnaire and NASA TLX.} We also asked her/him to provide feedback to improve the User Experience in OnToMap.
For the experiments we used a set of laptops with 15.6'' display and 1920x1080 resolution.

\section{Results}
\subsection{Demographic data and background}
For the user study, we recruited 62 participants (32.3\% women, 66.1\% men and 1.6\% not declared). Their age is between 20 to 70 years, with a mean value of 33.45.
They are part of the University staff (researchers, professors and secretaries) and students, as well as people working in the industry or retired.
In the pre-test questionnaire we analyzed their background and familiarity with technology:
41.9\% of participants have a scientific background, 29\% a technical one, 21\% humanities and linguistics, 6.5\% economics and law, 1.6\% arts.
Regarding the education level, 46.8\% of them attended the high school, 45.2\% the university, 6.5\% have a Ph.D and 1.6\% attended the middle school. 41.9\% of people declared that they use e-commerce platforms or online booking services monthly, 38.7\% said one or two times per year and 19.4\% weekly. Moreover, 56.9\% declared that they often use online services based on geographic maps, 17.7\% sometimes and 25.8\% every day.

\begin{table*}[t]
\centering
\small
\caption{Participants' performance during the execution of individual tasks. Time is expressed in seconds and the best values are in boldface. Significance is encoded as (**) $p<0.001$ and (*) $p<0.002$.}
\label{tab:performance}
\resizebox{\textwidth}{!}{%
\begin{tabular}{lcccc}
\toprule
Widget type & Min time & Max time & Mean time 
& Correct answers\\
\midrule   
                             
\multicolumn{1}{l}{
\renewcommand{\arraystretch}{0.7}
\begin{tabular}[l]{@{}l@{}} 1: Checkboxes \end{tabular}}                                                 & 33  & 184 & 94.26              
            & {\bf 100.00\%$^{*}$}  \\
\multicolumn{1}{l}{
\renewcommand{\arraystretch}{0.7}
\begin{tabular}[l]{@{}l@{}} 2: Treemaps \end{tabular}}                                                 & 33  & 180 & 77.39 
            & 98.39\%   \\
\multicolumn{1}{l}{
\renewcommand{\arraystretch}{0.7}
\begin{tabular}[l]{@{}l@{}} 3: Sunburst \end{tabular}}                                                     & {\bf 20} & 149 & {\bf 55.94}$^{**}$                 
            & {\bf 100.00\%$^{*}$}   \\
\multicolumn{1}{l}{
\renewcommand{\arraystretch}{0.7}
\begin{tabular}[l]{@{}l@{}} 4: Transparency sliders \end{tabular}}& 23 & {\bf 146} & 57.05                  
            & 95.16\%   \\               
\bottomrule
\end{tabular}%
}
\label{tab:timePerformance}
\end{table*}

\subsection{User performance}
\label{sec:performance}
Table \ref{tab:timePerformance} shows the results concerning participants' execution time and percentage of correct answers for each task. A Friedman test on execution times among the four tasks showed that there is a statistically significant difference between them:
$\chi^2(3)=207.57$, $p<0.001$, Kendall's $W=0.56$. The percentages of correct answers is statistically significant, too: $\chi^2(3)=14.14$, $p<0.002$, Kendall's $W=0.04$.

\begin{table*}[t]
\centering
\caption{Results of the post-task questionnaire. The best values are shown in boldface. Significance is encoded as (**) $p<0.001$ and ($\diamond$) $p<0.03$.}
\label{tab:task-questionnaire}
\small
\begin{tabular}{@{}lccccc@{}}
\toprule
Question \# & 1 & 2 & 3 & 4 & 5 \\ 
\midrule
\multicolumn{6}{c}{Task1: Checkboxes}                \\ \midrule
Mean        & \bf{3.90$^{**}$} & 4.03 & 3.77 & 4.02 & 2.94 \\
Variance    & 1.40 & 1.08 & 1.39 & 0.84 & 1.31 \\
St. Dev.    & 1.18 & 1.04 & 1.18 & 0.91 & 1.14 \\ \midrule
\multicolumn{6}{c}{Task2: Treemap}                    \\ \midrule
Mean        & 3.32 & 4.00 & \bf{3.95} & 3.98 & 3.48 \\
Variance    & 1.21 & 0.66 & 0.87 & 0.84 & 0.84 \\
St. Dev.    & 1.10 & 0.81 & 0.93 & 0.91 & 0.92 \\ \midrule
\multicolumn{6}{c}{Task3: Sunburst}                   \\ \midrule
Mean        & 2.95 & \bf{4.11} & 3.84 & 3.87 & \bf{4.10$^{**}$} \\
Variance    & 1.62 & 0.72 & 1.22 & 0.84 & 0.97 \\
St. Dev.    & 1.27 & 0.85 & 1.10 & 0.91 & 0.99 \\ \midrule
\multicolumn{6}{c}{Task4: Transparency sliders}       \\ \midrule
Mean        & 3.79 & 3.85 & 3.69 & \bf{4.31$^\diamond$} & 3.02 \\
Variance    & 1.28 & 1.21 & 1.20 & 0.87 & 1.52 \\
St. Dev.    & 1.13 & 1.10 & 1.10 & 0.93 & 1.23 \\ \bottomrule
\end{tabular}%

\end{table*}

\begin{table*}[t]
\centering
\small
\caption{Statistical significance of the post-task questionnaire results.}
\label{tab:friedman}
\begin{tabular}{@{}ccccc@{}}
\toprule
Question \# & Friedman's $\chi^2$ & df & $p$-value & Kendall's $W$ \\
\midrule
1           & 25.038              & 3  & $<0.001$    & 0.1346         \\
2           & 4.6779              & 3  & $=0.197$    & 0.0251        \\
3           & 1.4063              & 3  & $=0.704$   & 0.0076         \\
4           & 9.4442              & 3  & $<0.03$     & 0.0508        \\
5           & 43.611              & 3  & $<0.001$ & 0.2345         \\
\bottomrule
\end{tabular}%
\end{table*}

As shown in the table, people achieved the lowest mean execution time and they correctly answered 100\% of the questions when they used the widget based on the sunburst diagram. In comparison, when they used the checkboxes, they correctly answered all the questions but they spent the longest time to complete the task. By using the treemaps, participants spent a long time to perform the tasks (almost as long as with checkboxes) but they correctly answered 98.39\% of the questions. Finally, they spent relatively little time with transparency sliders but they provided 95.16\% correct answers. The high number of correct answers should not surprise because people could check them more than once. 

We observed that, in {\em Task1}, {\em Task2} and {\em Task3}, almost all the participants removed some irrelevant data categories using the transparency sliders to reduce map cluttering; then, they used faceted exploration to analyze data. 
However, they leaned to use the checkboxes embedded in the transparency sliders instead of using the sliders to tune the opacity of items.

\begin{table*}
\small
\centering
\caption{Post-test questionnaire (translated from the Italian language).}
\resizebox{\columnwidth}{!}{%
{\def\arraystretch{1.3}
\begin{tabular}{ll}
\toprule
\# & Question/statement \\ 
\midrule
1 & 
\multicolumn{1}{l}{
\renewcommand{\arraystretch}{0.7}
\begin{tabular}[l]{@{}l@{}}The widget was familiar to me.
\end{tabular}}
  \\ 
  
2 & 
\multicolumn{1}{l}{
\renewcommand{\arraystretch}{0.7}
\begin{tabular}[l]{@{}l@{}}The widget helped me find the information I needed.
\end{tabular}}
  \\ 
  
3 & 
\multicolumn{1}{l}{
\renewcommand{\arraystretch}{0.7}
\begin{tabular}[l]{@{}l@{}}The widget helped me to save effort in answering the questions.
\end{tabular}}
  \\ 
4 & 
\multicolumn{1}{l}{
\renewcommand{\arraystretch}{0.7}
\begin{tabular}[l]{@{}l@{}}The widget was easy to use.
\end{tabular}}
  \\ 

5 & 
\multicolumn{1}{l}{
\renewcommand{\arraystretch}{0.7}
\begin{tabular}[l]{@{}l@{}}The widget is novel.
\end{tabular}}
  \\ 

6 & 
\multicolumn{1}{l}{
\renewcommand{\arraystretch}{0.7}
\begin{tabular}[l]{@{}l@{}}Do you think that using transparency sliders in combination with checkboxes, \\ treemaps or sunburst diagram is useful?
\end{tabular}}
  \\ 
7 &\multicolumn{1}{l}{
\renewcommand{\arraystretch}{0.7}
\begin{tabular}[l]{@{}l@{}}Which information exploration widget would you use again in the future?\end{tabular}}
  \\
8 & 
\multicolumn{1}{l}{
\renewcommand{\arraystretch}{0.7}
\begin{tabular}[l]{@{}l@{}}Why? \end{tabular}}
  \\ 
9 & 
\multicolumn{1}{l}{
\renewcommand{\arraystretch}{0.7}
\begin{tabular}[l]{@{}l@{}}Which information exploration widget did you like the least?\end{tabular}}
   \\ 
10 &  
\multicolumn{1}{l}{
\renewcommand{\arraystretch}{0.7}
\begin{tabular}[l]{@{}l@{}}Why?  \end{tabular}}
 \\ 
 \bottomrule
\end{tabular}
}}
\label{tab:postTest2}
\end{table*}

\subsection{User Experience - post-task questionnaire}
Table \ref{tab:task-questionnaire} shows the results of questions 1-5 of the post-task questionnaire and Table \ref{tab:friedman} shows the results of a Friedman significance test applied to these results.
\begin{itemize}
    \item 
    {\bf Question 1 (familiarity):} participants were most familiar with the widgets based on checkboxes and, in second position, with the transparency sliders. They were less familiar with the treemaps and much less with the sunburst diagrams ($p<0.001$). 
    \item 
    {\bf Question 2 (helpfulness):} the results are not statistically significant but the generally high ratings prove that participants perceived all the widgets as helpful to find information items in the maps. The transparency sliders received the lowest ratings.
    \item 
    {\bf Question 3 (effort saving)}: the results are not statistically significant; however, similar to Question 2, the transparency sliders are evaluated worse than the other widgets. In this case, ratings show that participants felt that the widgets helped them to save efforts during task execution but values are a bit lower than those of Question 1. 
    \item 
    {\bf Question 4 (ease of use):} participants perceived transparency sliders as the easiest tool, followed by the checkboxes, treemaps and sunburst diagram ($p<0.03$). This finding is in line with the results of Question 1: even though sliders and checkboxes are in a different preference order, the Pearson Correlation between the answers to Question 1 and Question 4 shows that they are positively correlated both on checkboxes ($\rho=0.5015$) and on transparency sliders ($\rho=0.4802$).
    \item 
    {\bf Question 5 (novelty):} participants perceived the widgets based on the treemaps and sunburst diagrams as more innovative than the other ones; they also evaluated the checkboxes as the least innovative one ($p<0.001$). 
    The Kendall's $W$ value (0.2345) is the best one across the five questions. This demonstrates that there is more agreement among participants about the perception of novelty of widgets with respect to the other evaluation dimensions, i,e., ease of use, and so forth.
    \end{itemize}
About a quarter of the participants answered the free text questions; the percentages reported below refer to this set of people. 
\begin{itemize}
\item
{\bf Question 6 (difficulties):}  50\% of the participants who answered this question declared that, due to the amount of textual information displayed in the checkboxes, they had difficulties in the identification of the widgets representing the categories of interest in the side bar. 
Some people pointed out that the treemap and the sunburst were new visualization models; thus, they initially had some difficulties in understanding how they worked.
A few participants complained about the shortening of facet values in the treemaps because they had to move the mouse over their components to read the information. The only observed limitation of the sunburst was that it is visualized in a separate window, partially covering the map.
\item 
    {\bf Question 7 (appreciations):} some participants liked the graphics of the treemaps and declared that the size of the components representing facet values provides an intuitive visualization of the cardinality of the corresponding sets of items.
        About 25\% of people perceived the sunburst as good to compactly visualize all the facets and values of a data category. They also appreciated the fact that the sunburst reduces the vertical expansion of the side bar; thus, it limits the scrolling to reach the widgets of interest. 
        Some participants specified that they liked the correspondence of colors between sliders and items in the map; i.e., color coding. 
        In general, the transparency slider was perceived as useful to reduce information overload by imposing visualization constraints on whole data categories.

\end{itemize}

 \begin{figure}[t]
 \centering
 \includegraphics[width=0.8\columnwidth]{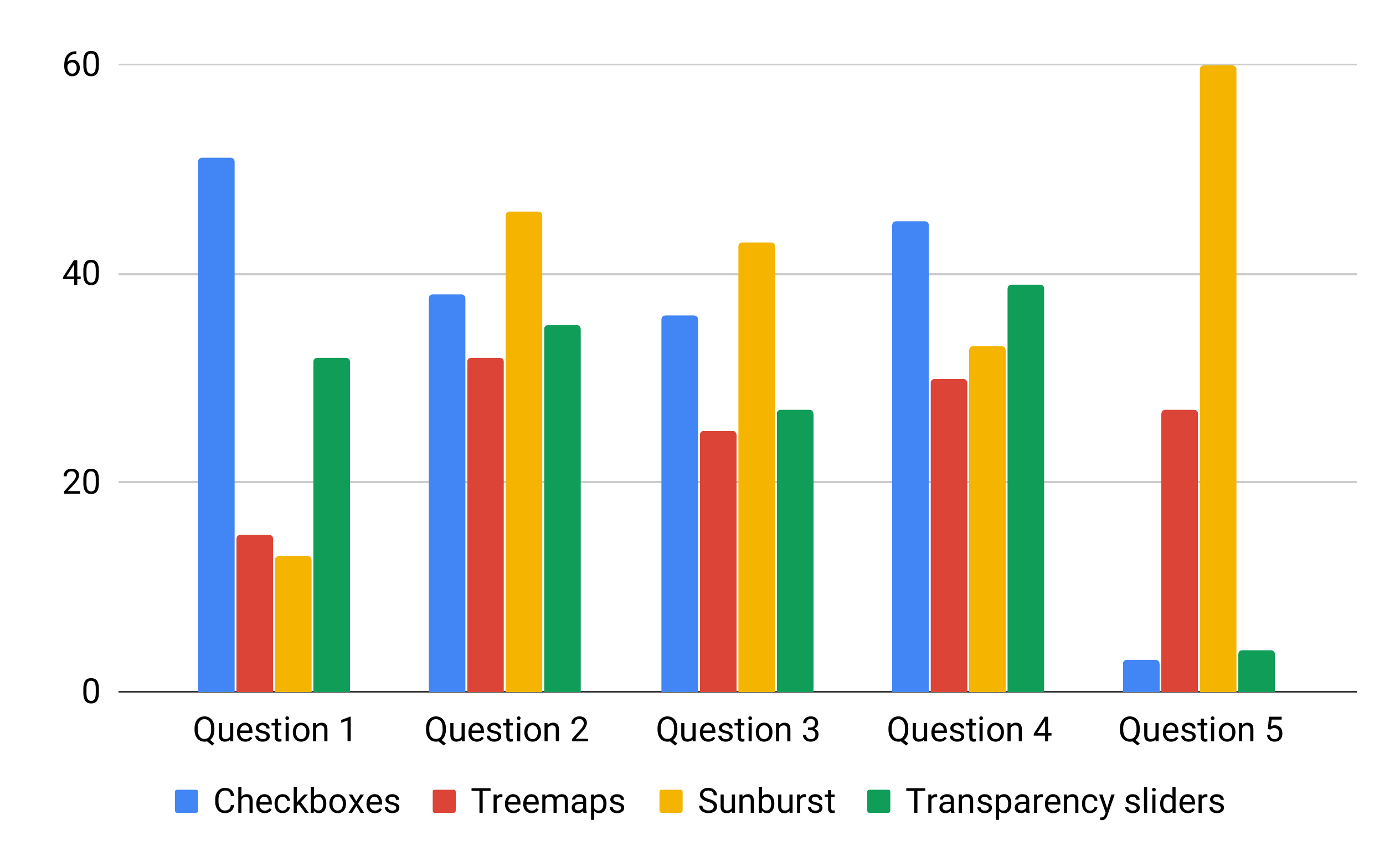}
 \caption{Post-test: evaluations of the questions listed in Table \ref{tab:questions}.}
 \label{fig:post-test}
  \end{figure}

\subsection{User Experience - post-test questionnaire}
After participants completed the four tasks, we asked them to fill in a post-test questionnaire to capture their overall experience with the widgets. 

In the first part of this test we asked them to select the widgets which better matched familiarity, helpfulness, effort saving, easy of use and novelty; see Table \ref{tab:postTest2}.
People could check multiple options in case more than one widget satisfied them; therefore, the percentages reported below may be over 100\%.
The results of this part of the test, shown in Figure \ref{fig:post-test}, are consistent with those of the post-task questionnaires. Specifically, they confirm that:

\begin{itemize}
    \item 
    {\bf Question 1 (familiarity):} people considered the checkboxes as the most familiar widget and they placed transparency sliders in second position.
    \item 
    {\bf Question 2 (helpfulness):} the sunburst was perceived as more helpful than the other widgets as far as information finding is concerned.
    \item 
    {\bf Question 3 (effort saving):} the sunburst, followed by the checkboxes, was the preferred widget from the viewpoint of effort saving. This is different from the results of the post-task questionnaires but it should be noticed that those results are not statistically significant.
    \item 
    {\bf Question 4 (ease of use):} the checkboxes were evaluated as the easiest widget to use, slightly easier than transparency sliders. In the post-task results were reversed but the two widgets were anyway the best rated ones.
    \item 
    {\bf Question 5 (novelty):} the sunburst was perceived as the most novel widget.
\end{itemize} 
 Regarding the second part of the post-test questionnaire (see Table \ref{tab:postTest2}):
\begin{itemize} 
\item
{\bf Question 6 (transparency sliders with facet-based widget)}: 53.2\% of participants declared that the joint usage of transparency sliders with checkboxes, treemaps or sunburst efficiently helps information exploration. They found it convenient to organize search in two steps: (i) visual simplification of maps by hiding the data categories irrelevant to the questions, using transparency sliders; (ii) identification of the items of the category of interest on the basis of their properties, using faceted exploration widgets.
\item 
{\bf Questions 7 and 8 (future usage of widgets and why)}: 60\% of people stated that they would use the widget based on the sunburst again because it offers a complete view of each data category. Moreover, 56\% declared that they would use the checkboxes again because this is a widespread way to search for information. 
\item
{\bf Questions 9 and 10 (least preferred widget and why):} 34\% of participants evaluated the treemaps as the least preferred widget because they are not intuitive and they are difficult to use; 
37\% did not like the transparency sliders either because they poorly help solving complex search tasks.
21\% of people did not like the sunburst, mostly because it covers part of the map instead of being displayed within the side bar. Finally, 15\% declared that they would not use the checkboxes in the future because they carry a large amount of textual information and it's difficult to identify the relevant values. 
\end{itemize}

\section{Discussion}
\label{sec:discussion}
The user performance and experience results consistently suggest that the sunburst is the best widget for faceted information exploration. Specifically, User Experience results can be explained as follows:
\begin{itemize}
    \item 
    Regarding the familiarity with the types of widget (Question 1), we expected that people would be more familiar with checkboxes and transparency sliders because they are used to support faceted search in several e-commerce and booking applications, while treemaps and sunburst are rarely used outside scientific contexts.
    \item
   Question 2 provides some evidence that participants perceived the widget based on the sunburst as the most helpful one (post-test), while transparency sliders were suitable to solve simple search problems because they do not support facet-based map projection (post-task and post-test). 
   People also considered the transparency slider as useful within a facet-based widget (Question 6 - post-test). 
    \item
    As far as effort saving is concerned (Question 3), the moderate appreciation and the mixed ratings given by participants might be explained by considering that, even though all the widgets support map projection, they require some interaction, which could be perceived as an effort. 
    \item
    Participants' familiarity with the widgets can explain the fact that they evaluated transparency sliders and checkboxes as the top easy-to-use tools (Question 4), and treemaps and sunburst as the most novel ones (Question 5). Moreover, the moderate ease of use attributed to treemaps and sunburst can partially depend on the fact that people had to learn how to use them (Question 6 - post-task).
\end{itemize}
Interestingly, in the answers to the free text questions (Question 6 - post-task and Questions 9 and 10 - post-test) a relevant number of participants criticized the amount of textual information visualized in the checkboxes, complaining that it challenges the identification of the relevant widgets or values in the side bar. Actually, all the faceted widgets include the same information, generated as described in Section \ref{sec:model}. Therefore, this comment can be interpreted in a different way, in relation to the lack of compactness of the layout provided by the checkboxes (and presumably also by the treemaps, even though they save a bit more space than the former).

The widget based on the treemaps was the least preferred one because it was not particularly intuitive and it was difficult to use (Questions 9 and 10 - post-test). Despite the appeal of its graphics (Question 7 - post-task), this widget challenges the user with readability issues. Moreover, similar to the checkboxes, it occupies a fairly relevant amount of vertical space in the side bar (Question 6 - post-task), thus increasing the amount of scrolling needed to inspect the other treemaps. 

We conclude that the experimental results help us answer our research questions, which we repeat here for the reader's convenience:
\begin{itemize}
    \item[RQ1:] 
   {\em Does faceted exploration of map content help users in finding the needed information in a geographic map that visualizes different types of data?}
    
    We compared participants' performance and experience using different widgets for faceted information exploration, with respect to transparency sliders alone. By using some of these widgets, people could complete a set of map learning tasks more quickly and/or precisely than by only using the sliders. However, participants appreciated the combination of faceted exploration with basic category hiding because the latter enables the user to quickly hide irrelevant data, thus reducing visual complexity in the map, and the former supports detailed exploration of relevant items. This is different from traditional faceted exploration, in which users search for information within a single category. It also indicates that, when maps show multiple types of information, faceted exploration is effective but can be strengthened by adopting a model that jointly supports coarse-grained and fine-grained data projection. 
    \item[RQ2:] 
    {\em How does a compact, graphical view of the exploration options available to the user, which also shows the status of the information visualization constraints applied to a map, impact on her/his efficiency and experience in data exploration?}
    
    The results of the experiment show that not all the facet-based widgets equally helped participants while executing the tasks of the experiment. The reason for this difference is in the capability of the widgets to clearly and compactly describe the search context. 
    
    Specifically, the widget based on the sunburst was considered as particularly useful and effective,  and it supported the best user performance. 
    This finding is in line with previous experiments; e.g., those by \cite{Stasko-etal:00}. It can be explained by the fact that the sunburst provides a compact representation of the facets of a category, supporting the readability of their values. The compactness of this widget also enhances the conciseness of the side bar; in turn, this reduces scrolling during faceted search. Conversely, the treemaps challenged participants because their graphical layout hampers readability and their vertical extension excessively increases the length of the side bar. 
    
    Participants appreciated both sunburst and checkboxes; however, when using the latter they completed the tasks slower than with the former. As the main difference between these two widgets is in their vertical extension (much more compact in the sunburst), we can say that this is the main dimension determining the difference in performance.
    
    The transparency sliders achieved the lowest user performance results because they fail to support the specification of fine-grained visualization constraints. However, using the sliders in combination with the other widgets was perceived as a very convenient approach because it enables to first focus the map on the categories of interest, and then further project it by imposing detailed visualization constraints on the remaining items.
    \item[RQ3:]
    {\em How much does the user's familiarity with the widgets for faceted exploration impact on her/his efficiency in search and on her/his appreciation of the exploration model they offer?}
    
    The results of the experiment suggest that the familiarity with the widgets does not influence users' efficiency in search: the best performance in task execution was achieved by using the sunburst, which most participants considered as moderately ease to use and they did not know before interacting with OnToMap.
    Moreover, familiarity positively influences people's disposition towards the faceted exploration widgets and their perception of ease of use; see the case of the checkboxes. However, participants appreciated the sunburst as well because it efficiently supports exploration at the expense of some initial learning effort. We can thus conclude that, if the widget is not too difficult to use, the functionality it provides can override the effect of its familiarity on user appreciation.
\end{itemize}

\section{Conclusions and future work}
\label{sec:conclusions}
We presented a faceted information exploration approach supporting a flexible visualization of heterogeneous geographic data. Our model provides a multi-category faceted projection of long-lasting geographic maps to answer temporary information needs; this is based on the proposal of efficient facets for data exploration in sparse and noisy datasets.
Moreover, the model provides a graphical representation of the search context by means of alternative types of widget that support interactive data visualization, faceted exploration, category-based information hiding and transparency of results at the same time.

We carried out a user study involving 62 people who have diverse familiarity with technology and with map-based online systems.
The results of this study show that, when working on maps populated with multiple data categories, our model outperforms simple category-based map projection and traditional faceted search tools such as checkboxes. Moreover, the layout that uses the sunburst diagram as a graphical widget supports the best user performance and experience, thanks to its clarity and visual compactness.
We thus conclude that this implementation is promising for flexible faceted exploration in Geographic Information Search.
The described work has limitations that we plan to address:
\begin{itemize}
    \item 
    Our model only supports the specification of hard visualization constraints on facet values; i.e., the items having a certain value of a facet are either shown, or hidden. However, the user might want to specify preferences. Therefore, similar to what has been done in some related works (see Section \ref{sec:related}), we plan to manage soft visualization constraints.
    \item 
    So far, we present search results in geographic maps and we provide item details in dynamically generated tables showing their properties. In order to enhance data interpretation and sensemaking, we plan to develop additional visualization models supporting visual analytics; e.g., see \citep{Andrienko-etal:07, Tsai-Brusilovsky:19,Cardoso-etal:19}.
    \item 
    We designed the questionnaires of our user study by taking inspiration from existing sources (NASA TLX and User Experience Questionnaire) but we personalized the questions in order to test the specific aspects which are the focus of the present paper. We plan a credibility/validity analysis to verify that our questionnaires are strictly related to these sources.
    \item 
    Further experiments are needed to validate the proposed model with a larger set of people and on mobile phones (the OnToMap user interface scales well to the screens of tablets). 
    \item 
    Currently, our model supports a ``one size fits all'' type of faceted search that exploits general efficiency criteria to guide the user in data exploration. However, some researchers propose to adapt facet suggestion to the user's preferences in order to personalize the navigation of the information space; e.g., see \citep{Tvarozek-etal:08,Tvarozek-Bielikova:10,Koren-Liu:08,Abel-etal:11}. In our future work, we plan to offer multiple data exploration strategies which the user can choose from, including a user-adaptive facet suggestion that depends on her/his preferences and on the search context.
    \item 
   Depending on their roles, in some scenarios users might need to access different, long-lasting custom views of a shared information space \citep{Rasmussen-Hertzum:13}. We plan to extend our model by introducing permanent, user-dependent views on map content.
\end{itemize}

\section{Acknowledgements}
This work was supported by the University of Torino which funded the conduct of the research and preparation of the article.

\newpage

\section*{Appendix}
\label{sec:appendix}

\begin{table*}[h]
\small
\centering
\caption{Entropy, mean cardinality and exploration cost of the facets displayed in Figure \ref{fig:comparison}.}
\begin{tabular}{lccc}
\toprule
                & Entropy & Mean Cardinality  & Exploration cost \\ \midrule
Outdoor Seating & 0.9416  & 46.0000 & 0.0205           \\
Takeaway        & 0.8482  & 25.3333 & 0.0335           \\
Ex1             & 3.0000  & 20.0000 & 0.1500           \\
Cuisine         & 4.3895  & 5.0233  & 0.8738           \\
Ex2             & 3.0000  & 3.0000  & 1.0000           \\
Name            & 9.3987  & 1.0000  & 9.3987          \\
\bottomrule
\end{tabular}
\label{tab:costdetails}
\end{table*}

\begin{table*}[h]
\small
\centering
\caption{Balance, object cardinality, frequency and navigation quality of the facets shown in Figure \ref{fig:comparison}, according to \citep{Oren-etal:06} with $\mu=2$ and $\sigma=4.9$. We remind that the colors of facets in Figure \ref{fig:comparison} correspond to the complement of the values reported in the present table.}
\begin{tabular}{lcccc}
\toprule
                & Balance & Object cardinality & Frequency & Navigation quality \\ \midrule
Outdoor Seating & 0.8587  & 0.9794             & 0.1280    & 0.1076             \\
Takeaway        & 0.5175  & 0.9201             & 0.1057    & 0.0503             \\
Ex1             & 1.0000  & 0.4725             & 0.2225    & 0.1051           \\
Cuisine         & 0.3663  & 4.54E-66           & 0.6008    & 9.99E-67           \\
Ex2             & 1.0000  & 0.4725             & 0.0334    & 0.0157           \\
Name            & 1.0000  & 0.0000             & 0.9388    & 0.0000            \\
\bottomrule
\end{tabular}
\label{tab:navqdetails}
\end{table*}





\end{document}